\documentclass[twocolumn,prl,aps,superscriptaddress]{revtex4-2}

\usepackage[colorlinks,linkcolor=blue,urlcolor=blue,citecolor=blue]{hyperref}

\usepackage{epsfig}
\usepackage{graphicx}
\usepackage{dcolumn}
\usepackage{bm} 
\usepackage{epstopdf}
\usepackage{amsmath}
\usepackage{color}
\usepackage{hyperref}
\usepackage{multirow}
\usepackage{threeparttable}
\usepackage{comment}
\usepackage{orcidlink}

\usepackage{booktabs}   

\newcommand{\beq}{\begin{equation}}
\newcommand{\eeq}{\end{equation}}
\newcommand{\beqa}{\begin{eqnarray}}
\newcommand{\eeqa}{\end{eqnarray}}

\allowdisplaybreaks[4]

\begin{document}

\title{Proton-proton Femtoscopy as a Probe of Short-range Structure in High-Energy O+O Collisions}

\newcommand{\FudanIMP}{Institute of Modern Physics, Fudan University, Shanghai 200433, China}
\newcommand{\KeyLab}{Key Laboratory of Nuclear Physics and Ion-beam Application(MOE), Fudan University, Shanghai 200433, China}
\newcommand{\ECNU}{School of Physics, East China Normal University, Shanghai 200062, China}

\newcommand{\Inner}{College of Physics and Electronics Information, Inner Mongolia Minzu University, Tongliao 028043, China}
\newcommand{\JoiLab}{Inner Mongolia Joint Key Laboratory of Nuclear and Radiation Detection, Tongliao 028043, China}
\newcommand{\lbnl}{Nuclear Science Division, Lawrence Berkeley National Laboratory, Berkeley, CA 94270, USA}

\author{\small Baoshan Xi}
    \affiliation{\Inner}
    \affiliation{\FudanIMP}
    \affiliation{\KeyLab}
    
\author{\small Pei Li}
    \affiliation{\FudanIMP}
    \affiliation{\KeyLab} 
   
\author{\small Chunjian Zhang}\email{chunjianzhang@fudan.edu.cn}
    \affiliation{\FudanIMP}
    \affiliation{\KeyLab}
\author{\small Jinhui Chen}
    \email{chenjinhui@fudan.edu.cn}
    \affiliation{\FudanIMP}
    \affiliation{\KeyLab}
    
\author{\small Su-Ya-La-Tu Zhang}
    \affiliation{\Inner}
\author{\small Yu-Gang Ma}
    \email{mayugang@fudan.edu.cn}
    \affiliation{\FudanIMP}
    \affiliation{\KeyLab}
    \affiliation{\ECNU}
\date{\today}

\begin{abstract}
Short-range nucleon-nucleon correlations are a defining feature of the nuclear many-body wave function, yet they are invisible in the one-body density and therefore inaccessible to observables that measure a nuclear size. We show that proton-proton femtoscopy supplies the missing sub-femtometer sensitivity. In $^{16}$O+$^{16}$O collisions at $\rm \sqrt{s_{NN}}=$ 200 GeV, we compare three nuclear-structure inputs spanning mean-field, low-resolution cluster, and short-range-correlated descriptions. The $p$-$p$ correlation function separates all three, most sharply in peripheral collisions, where the \textit{ab initio} input suppresses the extracted source radius by $\sim5\%$ relative to the mean-field baseline. Under identical conditions $\pi^{+}$-$\pi^{+}$ correlations respond an order of magnitude more weakly, and the $C_{pp}/C_{\pi^{+}\pi^{+}}$ double ratio retains the full effect, pointing to the short-distance weighting of the $^{1}S_{0}$ pair rather than to an overall rescaling of the source. The signal survives the leading theoretical systematic, the choice of strong-interaction potential, which we quantify explicitly. These results identify $p$-$p$ femtoscopy as a short-distance-resolved probe of light-nucleus structure, complementary to flow observables that constrain only the low-order moments of the initial geometry.

\end{abstract}
\maketitle

\textbf{Introduction.} Understanding nucleon-nucleon ($NN$) correlations from the strong nuclear force is a central goal of nuclear physics~\cite{Hen:2016kwk,Frankfurt:2008zv,Epelbaum:2008ga,VONOERTZEN200643,RevModPhys.90.035004,Cruz-Torres:2019fum}. While a mean-field description, e.g., a smooth three-parameter Fermi (3pF) density, captures the bulk structure of heavy nuclei, it breaks down for light and intermediate-mass systems, where many-body correlations become essential. Modern \emph{ab initio} approaches, rooted in chiral effective field theories of low-energy QCD, can now compute such correlations from first principles~\cite{Lynn:2019rdt,Bedaque:2002mn,Hammer:2019poc,Lee:2025req}. The $^{16}$O nucleus is an instructive case: its one-body density is close to spherical, yet the underlying many-body wave function exhibits a pronounced short-range repulsive hole in the two-nucleon distance distribution together with $\alpha$-cluster and tensor correlations, so that individual nucleon configurations fluctuate strongly event by event while the mean density remains featureless~\cite{Delaroche:2009fa,Giacalone:2024luz}. Distinguishing such correlations from a mean-field density is therefore not a matter of measuring a nuclear size, which requires resolving structure at a \emph{fixed sub-femtometer scale}, a task for which low-energy spectroscopy of the one-body density is intrinsically ill suited. 

Relativistic collisions of light nuclei provide a complementary route, translating the instantaneous nucleon configuration of the projectile into observable properties of the produced medium~\cite{Ollitrault:1992bk,Song:2017wtw,Shen:2020mgh,Summerfield:2021oex,Li:2025hae}. Anisotropic flow has been shown to retain memory of the initial geometry, including the enhanced short-range correlations present in \emph{ab initio} configurations~\cite{Giacalone:2024luz,Zhang:2024vkh,Giacalone:2024ixe,Li:2025bdn}. Flow and size fluctuations, however, respond to the low-order moments of the initial energy density---its eccentricity and its overall transverse size---and are therefore intrinsically coarse-grained: a short-range hole at $r\sim1$~fm enters only through its integrated effect on those moments~\cite{STAR:2024wgy,STAR:2025elk,Schenke,Giacalone:2024bud,Jia:2022ozr,Giacalone:2025vxa,Duguet:2025hwi}.

Femtoscopic two-particle correlations access a different projection of the same physics, namely the space-time separation distribution of the emitting source at freeze-out~\cite{Lisa:2005dd,Wiedemann:1999qn,Lednicky:2005af,Ollitrault:1997vz,Xi:2019vev,Fabbietti:2020bfg,STAR:2026ijb,Liu:2024uxn,He:2020jzd,Wang:2023ygv}. The projection is not automatically a fine-grained one. For identical bosons the correlation function is, up to final-state effects, the Fourier transform of the pair source, so its measurable low-$q$ region is controlled by $\langle r^{*2}\rangle$---a bulk scale that short-range correlations leave almost untouched~\cite{Pratt:1984su}. Explicit calculations bear this out: pion femtoscopic parameters respond to nuclear deformation, but not to $\alpha-$ clustering in $^{16}$O~\cite{Kincses:2025iaf}. Resolving a $1$~fm feature requires a channel whose two-particle weight is itself concentrated at $r^{*}\sim 1$~fm. Proton pairs are that channel. The $p$-$p$ system possesses a large negative $^1S_0$ scattering length ($a_{pp}\simeq-7.8$~fm) corresponding to a near-threshold virtual state, which produces a pronounced correlation maximum near $k^{*}\approx 20$~MeV/c whose height is controlled by the overlap of $|\Psi(r^{*},k^{*})|^{2}$ with the pair source at $r^{*}\lesssim 2$~fm, and which therefore scales inversely with the source rms radius~\cite{Koonin:1977fh,Sinyukov:1998fc,Stoks:1993tb,Tolos:2020aln,STAR:2015kha}. In short, $p$-$p$ femtoscopy thus weights the short-distance part of $S(r^{*})$ rather than its second moment. That this weighting can amplify small structural differences is supported by low-energy data, in which the proton emitting source responds an order of magnitude more strongly than the ground-state charge radius does~\cite{SRIT:2026qkv}, and by transport calculations in which switching on a short-range repulsive potential visibly modifies $C_{pp}$~\cite{Shen:2021dll}.

Extracting the absolute source radius from the $p$-$p$ correlation function requires the microscopic $NN$ interaction, and where the source is of order $1$~fm the overlap with the wave function is large enough that differences among strong-interaction potentials propagate directly into the fitted radius. The universality of the source term underlying this extraction has itself been questioned for strongly interacting pairs~\cite{Epelbaum:2025aan}. In practice, however, the approach has been shown to recover an
independently determined neutron-neutron scattering length from measured correlation functions~\cite{Si:2025eou}. Before $p$-$p$ femtoscopy can be used to constrain light-nucleus structure, the structural signal and this theoretical systematic must be placed on a common footing---a comparison that, to our knowledge, has not previously been made. Our conclusions therefore rest on ratios of radii extracted with a common potential, in which this ambiguity largely cancels, rather than on absolute values.

In this work, we quantify the sensitivity of $p$-$p$ femtoscopic correlations to initial nuclear configurations in $^{16}$O+$^{16}$O collisions at $\sqrt{s_{\rm NN}}=200$ GeV using a multiphase transport (AMPT) model, for three $^{16}$O inputs spanning mean-field, low-resolution cluster, and high-resolution short-range-correlated descriptions, and using $\pi^{+}$-$\pi^{+}$ correlations as a control channel. Our results show that peripheral collisions (60--80\% centrality) act as the premier experimental window, providing maximum discriminative power among initial nuclear configurations, with the extracted source radius shifting $\sim5\%$ while the corresponding $\pi^{+}$-$\pi^{+}$ effect stays below $0.5\%$, and this hierarchy follows the $r^{*}$-weighting of the two channels rather than the overall source scale. We further show that this structural sensitivity survives after explicitly controlling for the systematic uncertainty introduced by the choice of strong-interaction potential. These results establish $p$-$p$ femtoscopy as a quantitatively controlled probe of light-nucleus structure for upcoming precision measurements at RHIC and the LHC.

\textbf{Nuclear configurations and model setup.} Three initial configurations are chosen to span the resolution scale at which nucleon positions are specified, from none to $\sim1$~fm. 

The 3pF density provides an interaction-independent mean-field baseline~\cite{DeVries:1987atn},
\begin{equation}
\rho(r)=\rho_0\frac{1+w(\frac{r}{R})^2}{1+\exp\left(\frac{r-R}{a}\right)} ,
\label{eq:3pf}
\end{equation}
where $\rho_0$ is the normalization factor, the half-density radius $R=2.608$ fm, the surface diffuseness $a=0.513$ fm and the parameter $w=-0.051$~\cite{DeVries:1987atn}. Nucleons are sampled independently, so no two-nucleon correlation is present by construction.

The Variational Monte Carlo---Auxiliary Field Diffusion Monte Carlo (VMC for brevity) approach generates nucleon coordinate distributions by variationally minimizing the expectation value of an N$^2$LO chiral Hamiltonian, approximating the ground-state solution of the Schrödinger equation~\cite{Lonardoni:2018nob,Lim:2018huo}. It inherently captures short-range repulsion and tensor correlations arising from the spin-isospin structure of the nuclear force \cite{Lynn:2019rdt}. As a result, the two-particle distance distribution exhibits a short-range repulsive hole and may feature enhanced clustering or angular correlations, which significantly affect event-by-event initial-state geometric fluctuations \cite{Zhang:2024vkh}.

The pionless Nuclear Lattice Effective Field Theory (NLEFT) configuration employs low-resolution structural inputs~\cite{Bedaque:2002mn,Hammer:2019poc,Lee:2025req,PhysRevLett.119.222505}, where nucleon distributions are generated in relatively coarse lattices using a minimal Hamiltonian that includes only contact interactions. This framework applies at momenta well below the pion mass, where pionic degrees of freedom are integrated out and the nuclear force is described by contact operators and their derivatives~\cite{Summerfield:2021oex}. We take the configurations from that work, in which each lattice site carries a Gaussian smearing of width 0.84 fm, the proton charge radius. The smearing scale therefore sets the resolution of the input directly: long-wavelength cluster correlations are retained, while two-nucleon structure below $\sim$ 1 fm---precisely where the \textit{ab initio} density develops its repulsive hole---is absent by construction. NLEFT therefore sits between 3pF and VMC and serves as a control: any observable that responds to VMC but not to NLEFT is responding specifically to short-distance structure.

We implemented these three nuclear configurations into the string-melting AMPT model with version v2.26t9 and a partonic cross section of 3.0 $mb$~\cite{Lin:2004en, Lin:2021mdn} to simulate $^{16}$O+$^{16}$O collisions $\sqrt{s_{\rm NN}}=$ 200 GeV, where this model provides a reasonable description of RHIC data~\cite{PhysRevLett.127.242301,PhysRevLett.131.022301,PhysRevLett.128.022301,Zhang:2024vkh,Zhao:2024feh}. Following the RHIC-STAR convention, we select $|\eta| < 1$, $p_{\mathrm{T}}>0.2$ GeV/c and $p<10$ GeV/c for the correlated particles and define centrality by $N_{\rm ch}$ $(|\eta| < 0.5$, $p_{\mathrm{T}}>0.2$ GeV/c). The three configurations yield $\langle N_{\rm ch}\rangle$ agrees within each centrality class, consistent with the multiplicity measured by STAR~\cite{STAR:2025ivi}. The comparison is therefore made at effectively matched multiplicity, and the configuration dependence reported below cannot be attributed to a residual difference in particle production. 

\begin{figure*}[htbp]
    \centering
    \includegraphics[scale=0.59]{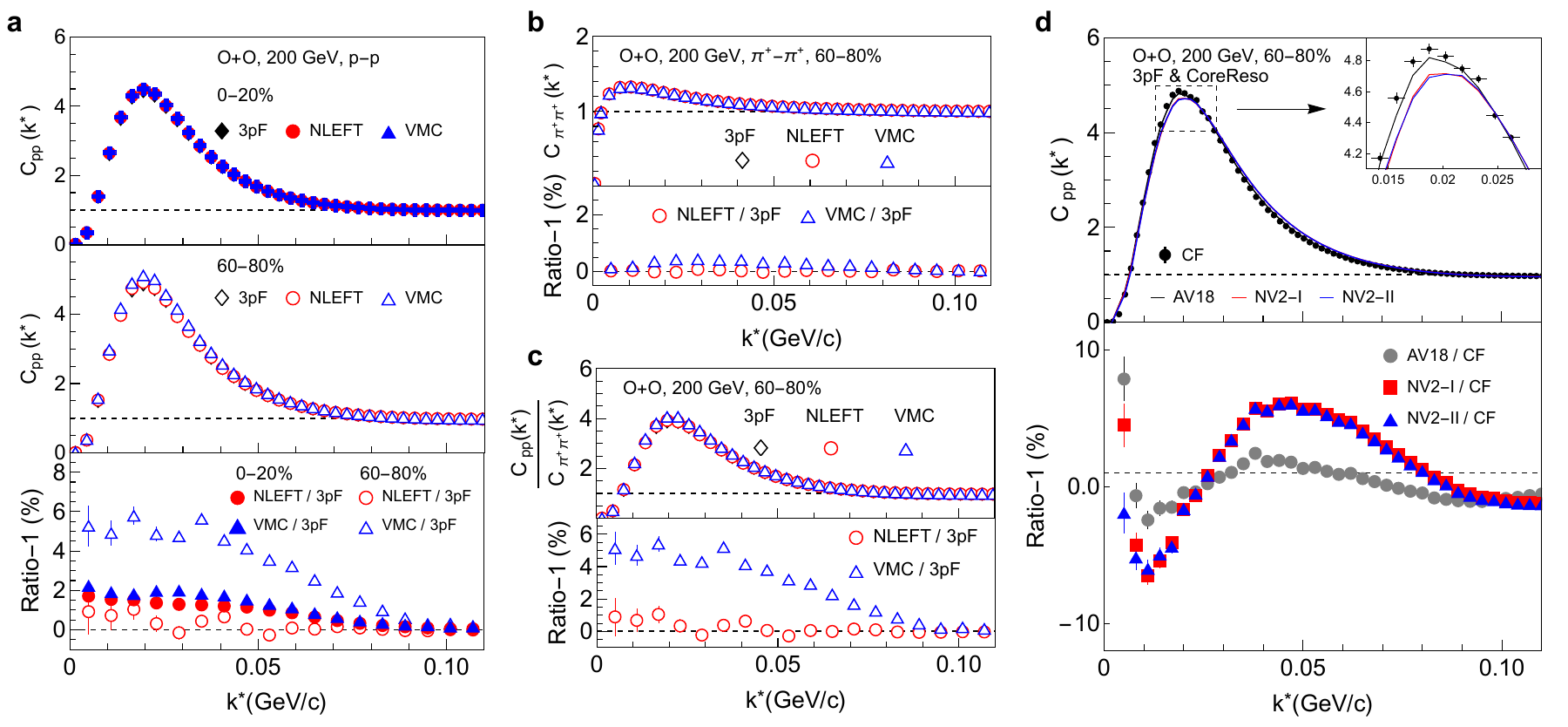}
    \caption{(a) Proton-proton correlation functions of O+O collisions of 0-20\% (upper panel) and 60\%-80\% (middle panel) for three different configurations and their ratios (lower panel). (b) Pion-Pion correlation function of O+O collisions of 60\%-80\% (upper panel) for three different configurations and their ratios (lower panel). (c) The ratio of $p$-$p$ correlations and $\pi^{+}$-$\pi^{+}$ correlations of O+O collisions of 60\%-80\% (upper panel) for three different configurations and their double ratios (lower panel). (d) Proton-proton correlation functions fitted with the 3pF model for the 60–80\% centrality, where the fit results correspond to AV18, NV2-I and NV2-II (upper panel). The ratios of the three fitted interaction curves to the $p$-$p$ correlation function are shown in the lower panel.}
    \label{Fig:1}
\end{figure*}

\textbf{Extracting correlation functions and source radii.} We evaluate the correlation functions with two complementary solvers. The Lednick\'y-Lyuboshits (LL) semi-analytical model provides a fast analytic description for a Gaussian source within the low-energy $S$-wave approximation~\cite{Lednicky:1981su,Lednicky:2005tb, Lednicky:1999xz, Lednicky:2008zz, Lednicky:2012zz, STAR:2014dcy,Chen:2024eaq,Wang:2024yke,Xi:2025iwd}. The Correlation Analysis Tool using the Schr\"odinger equation (CATS)~\cite{Mihaylov:2018rva}, by contrast, solves the two-body Schr\"odinger equation numerically for a given potential, allowing controlled inclusion of higher partial waves and systematic treatment of the signal region where the effective range expansion may break down, and has been adopted in many recent experimental femtoscopic extractions of hadron–hadron interactions~\cite{ALICE:2018ysd, ALICE:2019hdt, ALICE:2019gcn, ALICE:2020mfd, ALICE:2025byl}. Both start from the Koonin--Pratt relation~\cite{Boal:1990yh, Pratt:1990zq, ALICE:2025wuy},
\begin{equation}
C(k^*) = \sum_{}\int d^3\mathbf{r}^*\;
S(\mathbf{r}^*)\,
\Bigl|\Psi^{(+)}(\mathbf{r}^*,\mathbf{k}^*)\Bigr|^2,
\label{eq:kp}
\end{equation}
where $k^* \equiv |\mathbf{p}_1^*-\mathbf{p}_2^*|/2$ is the relative momentum in the pair rest frame, $S(\mathbf{r}^*)$ is the source distribution satisfying $\int d^3\mathbf r^{\ast}\,S(\mathbf r^{\ast})=1$, and $\Psi^{(+)}(\mathbf{r}^*,\mathbf{k}^*)$ denotes the two-body wave function for channel with outgoing boundary conditions. The strong interaction is provided by existing phenomenological potentials, AV18~\cite{Wiringa:1994wb} and ReidV8~\cite{Reid:1968sq}, or by tabulated coordinate-space potentials derived from the Norfolk NV2 chiral effective field theory family~\cite{Piarulli:2016vel}. For NV2, we employ coordinate-space partial-wave potential tables containing only strong interactions, while Coulomb interactions are treated separately within CATS to avoid double-counting electromagnetic contributions. The computational path from NV2 to CATS can be regarded as a more general treatment that replaces low-energy parameterization with potential models. Replacing a low-energy parameterization by explicit potential models maps the structural differences among $NN$ interactions directly onto shape differences of the correlation function---peak height, peak width, and asymmetric residuals in the signal region---and thereby provides the basis for the systematic comparison below.

Source radii are extracted by fitting the correlation functions within the $k^*$ over $0<k^*<0.15$ $\mathrm{GeV}/c$ under both Gaussian and core-resonance source assumptions. The Gaussian form corresponds to a spherically symmetric three-dimensional source,
\begin{equation}
S_{\rm G}(\mathbf r^{\ast};R_{\rm G})=
\frac{1}{(4\pi R_{\rm G}^{2})^{3/2}}
\exp\left[-\frac{r^{\ast 2}}{4R_{\rm G}^{2}}\right],
\label{eq:source_gauss}
\end{equation}
where $r^{\ast}=|\mathbf r^{\ast}|$ and $R_{\rm G}$ is the fitted Gaussian source radius with $\langle r^{\ast^2} \rangle_{\rm G} = 6R_{\rm G}^{2}$. In the core-resonance (CoreReso) model~\cite{ALICE:2020ibs,Becattini:2009ee}, primordial particles and short-lived resonances are emitted from a common Gaussian core $R_{\rm Core}$, while the propagation and decay of the resonances generate a non-Gaussian tail in source distribution, the pair separation being constructed as,
\begin{equation}
\mathbf r^* = \mathbf r_{\rm core}^* - \mathbf s_{{\rm res},1}^* + \mathbf s_{{\rm res},2}^*
\end{equation}
where $\mathbf r_{\rm core}^*$ is the primordial pair separation and $\mathbf s_{{\rm res},i}^*$ denotes the displacement of the $i$-th resonance before its decay, with resonance fractions and decay parameters following the ALICE experiment~\cite{ALICE:2020ibs}. A closure test with $\alpha$=2 and $f_{\rm res}=0$ confirms that the two implementations employ the same Gaussian-radius convention. Both parameterizations are therefore used to test the robustness of the extracted radii with respect to the assumed source shape, where their differences can be treated as part of the systematic source-model uncertainty.

\textbf{Results and Discussion.} \label{sec:result} Figure~\ref{Fig:1}(a) shows the $p$-$p$ correlation functions obtained with the LL
model for each configuration in the 0--20\% and 60--80\% centrality classes. The lower panel displays the relative ratios defined as $C(k^{*},\textit{ab initio})/C(k^{*}, \text{3pF})-1$ with NLEFT and VMC. The NLEFT/3pF ratio deviates by about $2\%$ in central collisions and is consistent with zero in peripheral collisions, whereas the VMC/3pF ratio grows from $\sim2\%$ in central collisions to $\sim5\%$ at 60--80\% centrality. Since NLEFT and VMC differ mainly in whether the $\sim1$~fm repulsive hole is resolved, the fact that only VMC departs from the baseline in peripheral collisions identifies short-distance structure as the origin of the effect. Overall nuclear size cannot account for it: although the three inputs differ in point-nucleon rms radius by up to 5\%~\cite{Zhang:2024vkh}, NLEFT has a smaller rms size than 3PF and yet tracks it, while the VMC--3pF separation \textit{grows} toward peripheral collisions, whereas a size difference would be most pronounced in central collisions, where all nucleons participate. The direction of the effect is specific to the high-energy regime. 

\begin{figure*}[htbp]
    \centering
    \includegraphics[width=\linewidth]{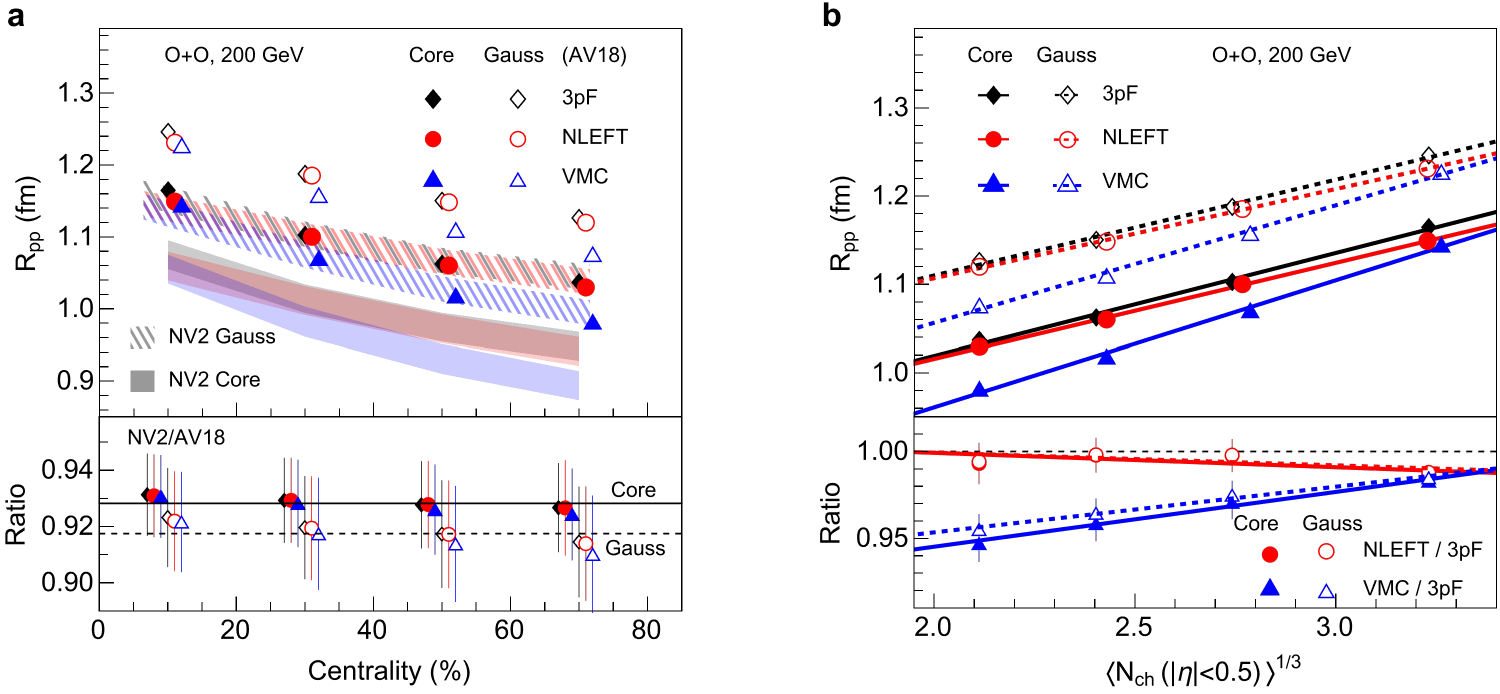}
    \caption{(a) Upper panel: the source radii are extracted for three cases under the AV18 potential using both Gaussian and CoreReso sources , while solid and hollow points correspond to the CoreReso and Gaussian sources, respectively. The maximum and minimum values obtained from four NV2 potential assumption are averaged to form the central values, with the differences serving as bandwidths. Lower panel: the ratio of source radii with NV2 and AV18 versus centrality. (b) Upper panel: the $\langle N_{\rm ch}\rangle^{1/3}$ corresponding to each centrality and used it as the horizontal axis show linear function fits. Solid and dashed lines for the Core and Gaussian results. Lower panel: the ratio fo source radii, $R_{\rm NLEFT}/R_{\rm 3pF}$ and $R_{\rm VMC}/R_{\rm 3pF}$. Each ratio point is plotted at the corresponding 3pF multiplicity coordinate.
    }
    \label{Fig:2}
\end{figure*}

It is important to emphasize that the observed configuration dependence does not arise solely from the $p$-$p$ final-state interaction. The 3pF, VMC, and NLEFT inputs first alter the initial nucleon spatial distribution of $^{16}$O; this propagates through event-level geometry and final-state expansion to reshape the freeze-out source function itself. $p$-$p$ femtoscopy resolves this reshaped source with high precision because its low-$k^*$ correlation is jointly governed by Coulomb repulsion, quantum statistics, and strong final state interaction. Other pairs offer complementary, if less clean, sensitivity. $\pi\pi$ and $KK$ trace the overall source scale and $m_{\rm T}$ scaling but are diluted by resonance-decay feed-down and late-stage rescattering; $p\Lambda$ and $\Lambda\Lambda$ might retain comparable sensitivity with larger interaction-model uncertainty; non-identical pairs such as $\pi K$ and $\pi p$ probe emission asymmetries instead. $p$-$p$ is thus a particularly sensitive, though not an exclusive, channel.

Figure~\ref{Fig:1}(b) presents the $\pi^{+}$-$\pi^{+}$ correlation functions for the $60$–$80\%$ centrality. In contrast to the $p$-$p$ channel, the $\pi^{+}$-$\pi^{+}$ ratios in the lower panel show that NLEFT/3pF is consistent with zero within uncertainties across the full $k^*$ range, while VMC/3pF exhibits a small residual enhancement of $<0.5\%$ at low $k^*$ that vanishes by $k^*\sim0.05$~GeV/$c$, an order of magnitude weaker than the $\sim5\%$ VMC/3pF deviation seen in $p$-$p$ (Fig.~\ref{Fig:1}(a)). This is expected behavior: identical-pion correlations at low relative momentum are governed by quantum statistics and Coulomb repulsion, whose kernel is sensitive to $\langle r^{*2}\rangle$ but flat over the sub-femtometer region where the configurations differ, and any residual sensitivity is further diluted by resonance feed-down and late-stage rescattering~\cite{Kincses:2025iaf,Lednicky:2005af,Lisa:2005dd}. The observability of nuclear-structure information in the final state is therefore strongly channel-dependent, with $\pi^{+}$-$\pi^{+}$ providing a weak but non-vanishing baseline rather than a null one.

This near-insensitivity makes $\pi^{+}$-$\pi^{+}$ a useful normalization baseline. Taking the double ratio $C_{pp}(k^*)/C_{\pi^+\pi^+}(k^*)$, shown in Fig.~\ref{Fig:1}(c), cancels common systematics associated with the overall source normalization while leaving the $p$-$p$-specific sensitivity intact. The double ratio for VMC/3pF remains at $\sim$5\% in peripheral collisions, essentially unchanged from the raw $p$-$p$ ratio in Fig.~\ref{Fig:1}(a). Had the $p$-$p$ signal originated in a uniform rescaling of the source, the double ratio would have been strongly
reduced; that it is consistent with that the sensitivity is carried by the short-distance weighting of the $^{1}S_{0}$ pair wave function.

Figure~\ref{Fig:1}(d) shows a representative CATS fit for the 3pF configuration in the 60--80\% centrality range under the CoreReso source, using AV18 and two NV2 variants, NV2-I (lpot=106) and NV2-II (lpot=110). The AV18 fit agrees well with the correlation function, while both NV2 fits are slightly poorer. The lower-panel ratios show AV18 closer to unity at low $k^*$, with NV2 systematically lower near the peak by $2$--$3\sigma$. This shape mismatch is confined to $k^* \lesssim 0.1$ GeV/$c$, where the low-energy partial-wave details of the potential are most influential and directly mapped onto the correlation peak via Koonin--Pratt convolution \cite{Koonin:1977fh, Lisa:2005dd, Mihaylov:2018rva}. The effective source radius extracted from the fit can adjust the overall peak amplitude, but cannot simultaneously eliminate the opposite-sign residuals on the two sides of the peak. At $k^* \gtrsim 0.1$ GeV/$c$, the correlation function approaches unity, the potential dependence fades, and the remaining fluctuations likely arise from baseline systematics, experimental resolution, or higher-order wave-function details \cite{Lisa:2005dd,Mihaylov:2018rva}.

The centrality dependence of the extracted source radii $R_{pp}$ from $p$-$p$ correlations is shown in Fig~\ref{Fig:2}(a), under Gaussian and CoreReso source assumptions using AV18 and NV2 potentials. Black, red, and blue points denote 3pF, NLEFT, and VMC, respectively, solid and hollow symbols correspond to CoreReso and Gaussian sources. Both enveloping bands are derived from the four NV2 variants. For all configurations and potentials, $R_{pp}$ decreases from central to peripheral collisions, consistent with typical femtoscopic behavior \cite{Lisa:2005dd}. The results of AV18 and Reid have been checked to nearly coincide across all centralities. The Gaussian radii $R_{G}$ are systematically larger than the CoreReso radii $R_{\rm Core}$, reflecting that non-Gaussian tails or long-lived components are absorbed into a larger effective radius in the single-Gaussian parameterization, while the CoreReso radius more closely represents the short-lived emission scale. The lower panel shows that the four NV2 variants produce a stable negative shift relative to AV18, $R_{\rm NV2}/R_{\rm AV18}\simeq0.93$ (CoreReso) and $0.92$ (Gaussian), with an internal envelope much narrower than the NV2--AV18 separation itself. This hierarchy implies that the potential-family choice dominates the systematics, while the NV2 variants reflect secondary effects from regularization scales and fitting windows \cite{Piarulli:2016vel, Mihaylov:2018rva}. Since Coulomb and quantum statistics are treated consistently in CATS, the NV2 offset is attributed to differences in the short-distance behaviour of the wave functions that modify the correlation strength after Koonin--Pratt convolution \cite{Koonin:1977fh, Mihaylov:2018rva}. The size of this shift provides a concrete measure of the scheme dependence recently emphasized for femtoscopic extractions involving nucleons~\cite{Epelbaum:2025aan}. Crucially, this offset is nearly independent of centrality, whereas the structural signal is not; the two are therefore separable in a measurement that compares centralities rather than absolute radii.

Figure~\ref{Fig:2}(b) shows the variations of $R_{\rm Gauss}$ and $R_{\rm Core}$ as functions of $\langle N_{\rm ch}\rangle^{1/3}$ for three nuclear configurations without pair-$m_T$ differentiation. If the particle density at freeze-out varies moderately, $N_{\rm ch}$ can approximately represent the effective freeze-out volume of the system; accordingly, $\langle N_{\rm ch}\rangle^{1/3}$ serves as a proxy for its characteristic linear scale. The central values of the slopes from six linear fits are all positive, demonstrating that the effective proton emission region expands overall with increasing collision multiplicity, in agreement with the trend reported by ALICE~\cite{ALICE:2025wuy}. The lower panel gives the radii relative to 3pF and linear function fits. The differences seen in the correlation functions propagate to the extracted radii: owing to the enhanced short-range correlations in VMC, the initial-geometry difference appears as a $\sim5\%$ reduction of $R_{pp}$ in peripheral collisions, while NLEFT stays consistent with 3pF at all multiplicities.

The peripheral enhancement of the signal has a simple physical origin, and we emphasize it because it determines where the measurement should be made. Three effects act in the same direction. First, the source is smallest in peripheral collisions ($R_{\rm Core}\simeq1$~fm), so the $^{1}S_{0}$ wave function overlaps the source precisely in the region where the configurations differ; the fractional response of the peak height to a change in radius, $d\ln C_{pp}/d\ln R$, grows as $R$ decreases. Second, peripheral events involve few participants and undergo little collective expansion and rescattering, so less of the initial two-nucleon information is erased before freeze-out. Third, the relative event-by-event fluctuation of the configuration is largest when few nucleons are sampled. Central collisions average over the full nucleus and wash out all three effects, which is why the VMC-3pF separation shrinks to the level of the NLEFT one there.

Taken together, these results provide a quantitative benchmark for how initial-state nuclear structure propagates into femtoscopic observables in a light-ion system. The peripheral sensitivity uncovered here also carries a practical message: experiments aiming to constrain nuclear structure through small-system collisions should
treat the initial geometry as a distinct systematic contri-


\textbf{Summary and outlook} We have investigated $p$-$p$ and $\pi^{+}$-$\pi^{+}$ femtoscopic correlations in $^{16}$O+$^{16}$O collisions at $\sqrt{s_{\rm{NN}}} = 200$ GeV with AMPT coupled to the LL and CATS frameworks, for three $^{16}$O inputs spanning mean-field, cluster-resolved and short-range-correlated descriptions. The $p$-$p$ correlations retain a genuine, configuration-dependent imprint of the initial nuclear structure: the VMC configuration, which encodes short-range nucleon-nucleon correlations, yields systematically smaller extracted source radii than the 3pF baseline, with the deviation growing from $\sim2\%$ in central collisions to $\sim5\%$ in peripheral collisions, while NLEFT stays close to 3pF throughout. This amplification pattern mirrors what has been established at low energy, where the proton emitting-source radius was found to respond an order of magnitude more strongly than ground-state charge radii do~\cite{SRIT:2026qkv}.

The hierarchy between channels follows from the resolution scale each provides. With its $^1S_0$ weight concentrated at $r^{*}\lesssim2$~fm, $C_{pp}$ resolves the short-range hole in the $ab$ $initio$ density. Identical-pion correlations, sensitive only to $\langle r^{*2}\rangle$, do not: their response stays below 0.5\%, an order of magnitude weaker, consistent with the null result reported for $^{16}$O clustering in pion femtoscopy~\cite{Kincses:2025iaf}, The resulting $C_{pp}/C_{\pi^+\pi^+}$ double ratio confirms that the nuclear-configuration sensitivity observed in $p$-$p$ is a genuine strong-interaction effect rather than an artifact of the overall source-size scaling shared by all particle pairs.

This sensitivity survives the competing systematic associated with the choice of strong-interaction potential, which we quantify explicitly. Relative to AV18, the Norfolk NV2 chiral-EFT potential introduces a stable radius shift $\Delta R \approx -0.09$ fm, with a distinct asymmetric residual shape in the $k^*\lesssim0.1$ GeV/$c$ signal region and only minor internal dispersion ($\sim0.02$ fm) among NV2 variants. Comparable in magnitude to the nuclear-structure signal itself, this shift quantifies, for the $p$-$p$
channel, the scheme dependence recently emphasized in
Ref.~\cite{Epelbaum:2025aan}; because it is nearly centrality independent while the structural signal is not, it cancels in the ratios on which our conclusions rest.

These results establish that $p$-$p$ femtoscopy offers a short-distance-resolved probe of light-nucleus structure, complementary to flow-based imaging, which constrains the low-order moments of the initial geometry instead. As high-precision light-ion data become available at RHIC and the LHC, the framework developed here provides a template for translating measured femtoscopic correlations into quantitative structural constraints.

\textbf{Acknowledgement.} We sincerely thank Jiangyong Jia, Dimitar Mihaylov, Yu Hu and Liang Zhang for the valuable discussions. This work is supported in part by the National Key Research and Development Program of China under Contract Nos. 2024YFA1612600 and 2022YFA1604900, the National Natural Science Foundation of China (NSFC) under Contract Nos. 12025501, 12547102, 12205051, the Natural Science Foundation of Shanghai under Contract No. 23JC1400200.

\bibliography{draft.bib}

\end{document}